# Application-driven engagement with universities, synergies with other funding agencies

Jim Hoff (jimhoff@fnal.gov)
Seda Memik (seda@northwestern.edu)

## Executive Summary

The seat of High-Energy Physics research in the United States is Fermilab.  However, considerable and essential research is also carried out at Argonne National Lab, Brookhaven National Lab, Berkeley National Lab and other federally funded laboratories as well as hundreds of universities nation-wide.  The interaction between the laboratories and the universities is, to use the colloquial phrase, *baked right into* the High-Energy Physics community.

The success of this Laboratory-University interaction is undeniable.  One need only look at the number of university professors and the number of senior scientists at the HEP Laboratories who were graduate student interns and fellows and then postdocs at the Laboratories before obtaining a tenure-track position or joining a national laboratory as a research scientist and advancing to the post they now hold.  Many senior scientists and academics owe much of their success to a system established decades ago that grew to become what is now called the High-Energy Physics community.

The success of the HEP Laboratory-University interaction is commendable. However, it is not inappropriate to point out that this Laboratory-University interaction has been historically limited in the United State to the physics departments of university partners. With the advent of automation, electronic instrumentation, and most notably data-driven scientific discovery in HEP, opportunities of collaboration between HEP laboratories and engineering departments, particularly those focusing on computation, electronics, and data sciences are becoming of utmost importance.  The accelerators and detectors and experiments in the HEP community represent one of the most significant, longest running and most fruitful engineering efforts undertaken in the history of this nation.  However, the synergistic relationship that exists between the HEP Laboratories and the university physics departments simply does not exist with engineering schools in the same widespread and systematic manner.  Interactions with engineering departments do happen occasionally, but, by and large, they are opportunistic. Unlike our colleagues in Europe, the United State High-Energy Physics community lacks the programmatic ability to sponsor engineering research at our universities, to place graduate students on engineering projects long-term, to request research into areas of significance to projects and to be able to influence thesis and dissertation topics.

In this whitepaper, we argue that nurturing HEP Lab-Engineering cooperation through established collaboration and support mechanisms will advance the scientific mission of the labs significantly, while at the same time giving the laboratories a stronger position in influencing the next generation workforce of engineers that will provide their services towards the unique computing and technology needs of the HEP community. The authors of this whitepaper are electronics and computer engineers and so, naturally, the arguments herein are made from their perspective.  However, these arguments are only strengthened by the simple fact that they could also have been made from the perspective of mechanical engineers, civil engineers or numerous other technologists. At the same time, this document serves as a summary of discussions that occurred during the Joint Instrumentation Frontier & Community Engagement Frontier Townhall meeting on November 10, 2020 (attended by Farah Fahim, Louise Suter, James Hoff - Fermilab, Helmut Marsiske - DoE, Seda Memik – Northwestern University, Ping Gui – Southern Methodist University, Valerio Re – Universita di Bergamo, Luca Carloni – Columbia University, Matt Shaw – Jet Propulsion Laboratory, and Sachin Junnarkar – Fieldviewers).

# 1. Introduction

Two important trends are becoming increasingly prominent in scientific projects within the HEP community as well as other scientific fields. The first, and historically preceding, development is the growing relevance of system optimization and integration methodologies in experimental systems. Sensor, control, and analysis tasks are getting integrated into high-speed, high-complexity electronic systems, often with increasing autonomous functionality. Design and deployment of these embedded components within the experimental systems require a combination of engineering expertise ranging from interface design, energy management, and data communication, to advanced integrated circuit technologies for miniaturization. The second trend concerns the proliferation of data driven approaches to scientific discovery. Many HEP scientists are relying heavily on data analytics to extract knowledge from the results of their experiments. This final extraction stage increasingly requires Artificial Intelligence/Machine Learning (AI/ML)-based methodologies. While entry-level practical applications of the AI/ML methods are becoming accessible to domain scientists of all fields, their deployment within sophisticated HEP experiments requires careful co-design of measurement systems and data processing from the bottom up. This, in turn, requires expertise in not only the practice of AI/ML, but also its theory, which is an active field of study in computer science and computer engineering departments of universities. Furthermore, many advanced HEP projects require these methods to execute in real-time systems, intimately embedded at the front-ends of the experiments. This unique technical requirement draws the engineering expertise into the mix, where electrical and electronic engineering expertise in hardware design needs to collaborate with computing experts and HEP scientists to design and deploy next generation experimental systems.

# 2. Mechanisms for Engagement

Considering the emerging trends discussed in the earlier section, we argue that the broader science and engineering community of HEP Labs and universities need to closely examine our track record, take notice of best practices, and formulate processes and workflows for optimal synergy moving forward. We identify three main directions as preliminary steps towards generating repeatable and sustainable arrangements.

2.1. Application-driven Project Partnership

Most productive partnerships start out with entities joining forces to achieve outcomes targeting a concrete problem. In that regard, application-driven engagement of HEP labs with universities will serve as a starting point. Ultimately, the intersection of the research interests between an HEP lab and a university should encompass a problem that the HEP lab considers of strategic importance. Fortunately, due to the reasons we have elaborated in the Introduction, there are fundamental engineering challenges that are expected to occur in a large variety of future HEP projects. Many involve intimate integration of measurement and on-demand computation via advanced integrated circuit technologies. Equally many will generate a deluge of data in need of real-time processing with intelligent and autonomous agents. Viewing these major technological parameters as pillars of HEP activities, it is easy to define a sustainable overarching technical synergy in the form of hardware/system/computation co-design that is bound to remain relevant within the limits of not just a one-off project, but a whole generation of new HEP endeavors.

A few concrete steps might further help to start define a landscape that is easier to navigate by the HEP scientists and university engineering departments. Some of those suggested mechanisms, such as workforce development and creating mutual positions will be described in greater detail in the later subsection.  As far as formalizing project domains are concerned, we would benefit from more explicit representation of these partnerships in established listings of funding opportunities as a starter. It would help both sides of the communities to have access to visible and specialized partnership support that is labeled clearly as HEP-Engineering effort. For instance, there are special fellowships reserved for graduate students from Physics to join an HEP lab as a yearlong fellow, while such specialty programs do not exist for Engineering graduate students in an exclusive manner. As for funded (externally or from within a lab's budget) research projects, we may want to consider a class of projects designated as science-engineering partnership, where the expectation of all project application in that category could be to include two dedicated components reserved for science and engineering, respectively.  In current practice, many lab-university joint projects proposed for funding to DoE or NSF certainly contain aspects of both domains, yet their co-presence is not something we acknowledge explicitly. Dedicating a track of projects where these components are required (similar to NSF projects where a research and an education component are requested to be described individually) would enhance the awareness in the community on the relative importance and value of the engineering contributions.

2.2. Workforce Training

Universities are the institutions where the US workforce for science and engineering is produced. A strong and competitive workforce stream with predictive supply trends is of utmost national value. HEP labs can make a conscious effort to becoming influential on the training of the next generation technologists through a number of mechanisms. These steps would in fact not only impact graduate level trainees, but also undergraduate students:

- All competitive research universities have strong traditions of involving undergraduate students in research. HEP labs can have "first dibs" on these students through systematic mechanisms. We have observed many individual success stories of HEP lab members mentoring undergraduate students. Some examples will be described in Section 3. A possible pipeline could be as follows. HEP labs recruit undergrads for internships through partner universities –> students are trained in the setups and topics the HEP lab prioritizes –> successful undergraduate students are channeled to graduate programs across a network of partner universities –> universities recruit these into engineering PhD programs –> co-advising models are used to mentor these students by both an engineering professor and a HEP scientist –> feed the students back into the HEP workforce.  [As a side note, the significance of a steady stream of HEP-experienced technologists who chose not to enter the HEP workforce, but instead join the general workforce should not be overlooked.  Research projects frequently require the

- collaboration of industry partners and HEP-trained engineers in the general workforce provide natural inroads for collaboration.]
- Another component of the workforce is international students who join US universities for graduate studies. Universities should welcome input from HEP labs on recruiting the next generation graduate students from this cohort with interdisciplinary (physics, science, and engineering) background.
- A practical mechanism for involving HEP scientists in recruiting and mentoring efforts of the universities is through establishment of joint academic appointments. All research universities have some form of Adjunct position defined for esteemed members of government labs and industry, so that they can have a direct link to provide input and interact with academics. Such positions should be reserved within engineering department for more HEP community members. These HEP scientists could then also be thesis advisors and thesis committee members of students. We must note that there are existing mechanisms to include external members on thesis committees. However, being an Adjunct faculty provides more direct access to university resources and information delivered on a daily basis to other faculty.
- Finally, active advocacy by HEP lab members for engineering students for awards, such as the URA award will be invaluable. Often times, engineering faculty and engineering PhD students are not aware of all opportunities that exist within the HEP and national lab ecosystems. Guidance from HEP scientists will help lift entry barriers for them.

## 3. Impediments to Collaboration

The Snowmass Townhall: Joint Instrumentation Frontier & Community Engagement included a discussion with Helmut Marsiske of the Department of Energy's Office of Science. He pointed out that Quantum Computation and Sensing may be the only research areas where multi-agency collaboration is accepted from the top down. Other research areas do not share this benefit. There is no single, central "pot of money" to draw from and, in fact, it is even difficult to share funding across departments within the DOE. Therefore, Lab-University collaborations that attempt to call upon multiple agencies to fund the same research will be met with resistance. Moreover, affinities or technical biases of grant reviewers and program managers, which inadvertently impose barriers, discourage them from allowing resources allocated within a certain domain to go "out of their field." For example, if an HEP Lab-university collaboration sought money from the Office of High Energy Physics to perform research into radiation tolerant device design, but the university collaborators were also pursuing similar research for NASA, it might be seen by the HEP program manager as giving HEP money to NASA research.

It is therefore necessary at this time to convince program managers that these types of collaborations are not only beneficial but required. To continue the analogy, both HEP and NASA require radiation tolerant electronics research. In fact, they require the same research to be done. It is to be expected that university experts in this field will attract the attention of both HEP and NASA, and there really is no reason to do the same research twice. Therefore, it should be clarified to the program managers that a joint HEP-NASA-university collaboration is beneficial to all even if it means that funding from different agencies must be mixed. This type of collaboration must happen, according to Dr. Marsiske, from the bottom-up until its significance is universally accepted and it becomes, like Quantum research, funded by multiple agencies top-down.

# 4. Examples of Successful Collaboration

One of the more successful, long-term collaborations between an HEP Laboratory and a university electrical engineering department has been between Fermilab and Dr. Ping Gui's Integrated Circuits and Systems Lab at Southern Methodist University. Significantly, Dr. Gui's group came to the attention of Fermilab's Electrical Engineering Department/ASIC Group through the reputation that her lab gained by their work on CERN's lpGBT (low-power Gigabit Transceiver) project. CERN has maintained a commitment to engineering research (as opposed to physics research) as a means to advance and facilitate physics experiments, even those that could be a decade or more in the future. lpGBT is a multi-year, multi-institution and, in fact, multi-national project spearheaded and funded by CERN that has developed a radiation-tolerant, configurable, high-speed data link for any HEP experiment. It is noteworthy that the lpGBT is now a fundamental piece of virtually every experiment under development in the HEP community and that the lpGBT itself is responsible for dozens if not hundreds of master's theses and doctoral dissertations. Dr. Gui and her team were responsible for much of the line driver, vcsel driver, TIA, and timing circuit developments necessary for lpGBT.

The collaboration has been very fruitful. It was responsible for some of the fundamental device research into the cryogenic lifetime and reliability of 65nm and 130nm CMOS technologies. It developed the design rule requirements for cryogenic ASICs. It has produced ten integrated circuits (including prototypes). It has resulted in a number of significant papers, several master's theses and doctoral dissertations. Fermilab has hosted many of Dr. Gui's students at the Lab, some for only a few days, some for weeks and months at a time. Most significant of all, one student has accepted a job as a member of the technical staff at Fermilab. This student, now colleague, was required to give a lecture at Fermilab and subject himself to the interview process as would be expected of any applicant to the Lab, but the simple truth is that his interview lasted several years. Fermilab did not extrapolate his design and lab skills from his resume. Fermilab engineers observed him in the Lab. His on-boarding was simplicity itself because he already knew the staff and he already knew his way around the Lab.

Another successful collaboration has been established between Fermilab and Dr. Seda Memik's lab at the ECE Department of Northwestern University. The Northwestern team has contributed to the first prototype chip for the Endcap Timing Readout Chip developed for the CMS Endcap at HL-LHC and they were involved in the late stage power and thermal modeling for the VIPRAM2D ASIC as part of a CMS L1 Tracking Trigger demonstration. Two PhD theses resulted from these projects. One of these students was the first ever engineering student recipient of the Universities Research Association (URA) Visiting Scholars Fellowship from Northwestern and his nomination was only possible because of strong support from Fermilab scientists. More recently, Dr. Memik's lab has been contributing to several machine learning accelerator ASIC chip designs that are part of the efforts to design a range of front-end data compression components for the HL-LHC and also they are the university partner of a DoE funded project led by the Fermilab on FPGA-based ML systems to infer and control beam loss de-blending for the Fermilab Main Injector and Recycler. Dr. Memik's students perform ASIC/FPGA design and verification tasks for a variety of computational elements in these systems. Furthermore, they are helping Fermilab ASIC design teams in setting up automated and systematic Computer Aided Design flows, which can be reused and re-adapted to many other future chip design tasks regardless of the specific experiment or front-end system configuration.